# SOLVING 3D RADAR IMAGING INVERSE PROBLEMS WITH A MULTI-COGNITION TASK-ORIENTED FRAMEWORK

*Xu Zhan, Xiaoling Zhang, Mou Wang, Jun Shi, Shunjun Wei, Tianjiao Zeng*

University of Electronic Science and Technology of China

## ABSTRACT

This work focuses on 3D Radar imaging inverse problems. Current methods obtain undifferentiated results that suffer task-depended information retrieval loss and thus don't meet the task's specific demands well. For example, biased scattering energy may be acceptable for screen imaging but not for scattering diagnosis. To address this issue, we propose a new task-oriented imaging framework. The imaging principle is task-oriented through an analysis phase to obtain task's demands. The imaging model is multi-cognition regularized to embed and fulfill demands. The imaging method is designed to be generalized, where couplings between cognitions are decoupled and solved individually with approximation and variable-splitting techniques. Tasks include scattering diagnosis, person screen imaging, and parcel screening imaging are given as examples. Experiments on data from two systems indicate that the proposed framework outperforms the current ones in task-depended information retrieval.

## 1. INTRODUCTION

3D radar imaging technology has shown potential values in multiple areas like urban structures' health [1], [2], military equipment scattering diagnosis [3], [4], person screen imaging [5], and parcel screening imaging [6], [7], etc. Show in Fig. 1, by transmitting and receiving a wideband signal at different positions, the target's 3D image can be solved.

Imaging is a typical inverse problem. The most applied methods, like range migration (RMA) [8] and back-projection (BPA) [9], belong to the matched-filtering (MF) imaging framework. Unavoidable side effects exist, like surrounding clutter and noises. To suppress them, researchers differentiate the target by inducing the sparsity feature [10]–[13], forming a sparse-oriented imaging framework, also known as sparse imaging in the area [14].

However, it may still be insufficient. Considering different task has variable demands, results are coarse and still undifferentiated. For example, the scattering diagnosis task for locating abnormal scatters cares about energy estimation accuracy. However, tasks like screen imaging don't care much. On the other hand, shape estimation matters more to detect and recognize dangerous objects. Such detailed differences have been ignored before, causing task-depended information retrieval loss.

To address this issue, a multi-cognition task-oriented framework is proposed. Aiming at better task-depended information retrieval, aspects including the imaging principle, model, and method are considered. Three aspects correspond to task needs: analyzing, embedding, and fulfilling. Specifically, the imaging principle is modified from MF/sparse-oriented to task-oriented, where an additional phase is introduced to analyze the task's demands. And the imaging model is modified from sparse-feature regularized into multi-cognition regularized, where every demand is embedded through a corresponding cognition regularization. To avoid couplings hindering the generalization, approximation [15], [16] and variable splitting [17], [18] techniques are introduced. After decoupling, different demands are fulfilled through solving sub-problems by proximal operators with gradient descent [19]. Three typical tasks are given as examples, including scattering diagnosis, person screen imaging, and parcel screen imaging. We conduct experiments on data from two different 3D radar systems. And they cover both the microwave band and the millimeter band. Results verify the proposed framework's generalization and superiority in task-depended information retrieval.

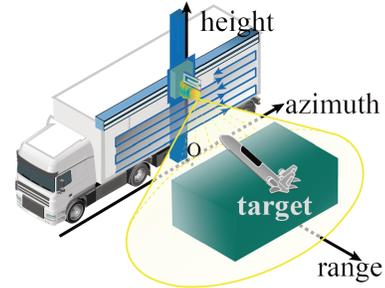

**Fig.1.** 3D radar imaging system model (ground-based array SAR).

## 2. MULTI-COGNITION TASK-ORIENTED FRAMEWORK

The forward process from the image $\alpha_{x,y,z}$ to the demodulated echoes $y_{x_p,y_p,r}$ can be expressed as

$$y_{x_p,y_p,r} = \sum_{x,y,z} \alpha_{x,y,z} s_t(r - r_{x,y,z}) + n_{x_p,y_p,r} \quad (1)$$

where $x_p$, $y_p$ and $r$ are sampling indexes for three directions, $x, y$, and $z$ are image indexes. $s_t(r - r_{x,y,z})$ is the $r_{x,y,z}$ delayed transmitting signal. $n_{x_p,y_p,r}$ is the noise-and-clutter.

In practice, only limited bandwidth echoes can be measured, which undermines the backward/inverse process, and only partial information can be retrieved. In this work, we focus on retrieving target-depended information. A new multi-cognition task-oriented framework is proposed, as illustrated in Fig.2. It's designed with considerations through the whole process from three articulated aspects, including the imaging principle, the imaging model, and the imaging method.

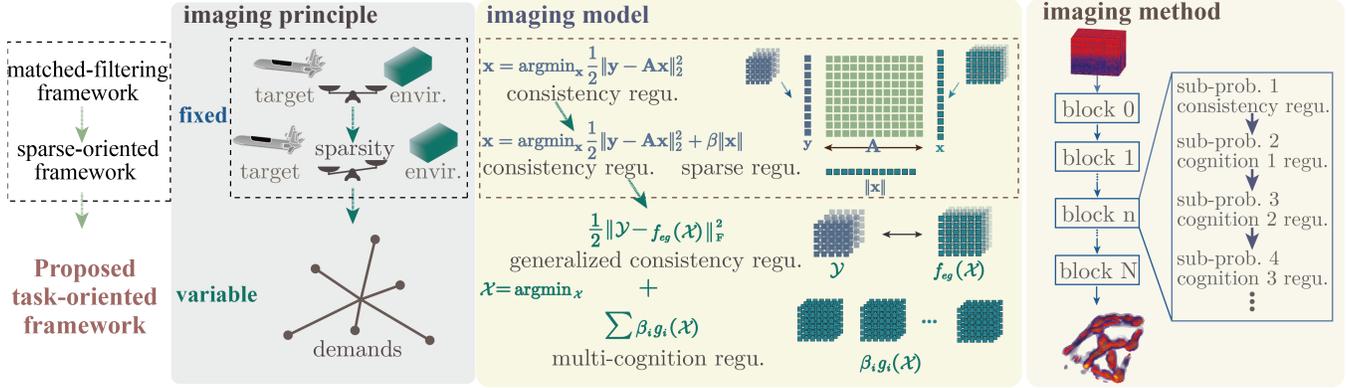

**Fig.2.** Proposed framework and the comparisons with current ones. Envir. is for environment, and Regu. is for regularization.

## 2.1. Task-oriented imaging principle

The most applied framework is the matched-filtering one, where the target and the environment are treated equally. To differentiate the target from the environment, the sparse-oriented one is proposed. However, still, only the difference between the target and the environment is considered. These two ones are fixed no matter what the task is. As the radar system gradually develops to measure larger bandwidth, a higher resolution image is available that contains more details. Thus, an opportunity is revealed for us to pursue a result that meets differentiated demands from tasks, as shown in the left column in Fig.2. We call it the task-oriented principle.

The most distinct feature is an additional analysis phase compared to current ones. The task is analyzed to obtain specific demands beyond the target's and the environment's aspects. As different tasks may prefer different demands, the resulting demands are variable in terms of the number and the content. Here, three tasks are given as examples.

1) *Scattering diagnosis*. This task is to locate the main point-scatters on the target, and then decide if they are normal or not according to the energies and positions in the image [20]. Thus, it has two demands. First, main scatters are retrieved with accurate amplitudes and positions. Second, clutter and noise from the environment are suppressed, avoiding forming false alarms. 2) *Person screen imaging*. Unlike the last one, the goal is to detect and recognize concealed dangerous objects on the human body [21]. It highly relies on the shape of the target. Thus, not only do the main point-scatters need to be retrieved, but the relatively weak distributed-scatters, which make up the shape, are also needed. The first demand is to retrieve main point-scatters, and the second is to retrieve distributed-scatters. 3) *Parcel screen imaging*. Although this task also aims at dangerous objects, different from the personal screening one, it is faced with a more complex environment. The target is placed inside a surrounding box or bag that can form a strong interference image [22]. Thus, the third additional demand is required to suppress the interferences from the surroundings.

## 2.2. Multi-cognition regularized imaging model

For simplicity and generalization, (1) can be expressed in the matrix-vector form, as shown in the middle column in Fig.2. Commonly, a consistency regularization term exists that guarantees measured echoes and estimated scatters are inconsistent with the electromagnetic wave propagation rule. In the model of the sparse-oriented framework, the target is enhanced through an additional sparse regularization term, which is mathematically described by the target's $l_1$ norm [10].

As multiple demands exist for a task, each corresponds to a cognition term. Intuitively, we can extend the sparse-oriented model into one with multi-cognition regularization terms.

$$\mathbf{x} = \text{argmin}_{\mathbf{x}} \frac{1}{2} \|\mathbf{y} - \mathbf{A}\mathbf{x}\|_2^2 + \sum_i \beta_i g_i(\mathbf{x}) \quad (2)$$

Where $\mathbf{x}$ and $\mathbf{y}$ are the folded target image and echo image, respectively. $g_i(\mathbf{x})$ is the $i$th cognition regularization term, and $\beta_i$ is the weight. These terms are mathematical descriptions to fulfill the different demands. The following sub-section will give examples corresponding to the former three tasks.

## 2.3. Generalized imaging method

The model in (2) seems to be reasonable at first glance. However, three critical insights should be pointed out. First, as the target and echo images have to be reshaped into 1D vectors to fit the model, their original 3D tensor structures are destroyed. This structure integrity loss may have little influence on the sparsity regularization because it does not describe the mutual information of pixels. However, to fulfill multi-cognition regularizations, this loss can't be ignored. For example, shape cognition usually relies on the original 3D space. Second, to fit the folded x, the observation matrix A would be relatively large, causing large storage and computation burdens [12]. Third, there are obvious couplings between regularization terms. As these regularization terms are variable for the task-oriented imaging framework, the coupling problem can hinder the generalization of the method.

To address the first and second issues, we introduce the approximation technique [15], on the basis of (2), we modify it as follows.

$$\mathcal{X} = \arg\min_{\mathcal{X}} \frac{1}{2} \|\mathcal{Y} - \text{f}_{\text{eg}}(\mathcal{X})\|_F^2 + \sum_i \beta_i g_i(\mathcal{X}) \quad (3)$$

Where $\mathcal{X}$ and $\mathcal{Y}$ are original 3D tensor forms of target image and echo image, respectively. And $f_{eg}(\cdot)$ is a generalized version of the observation matrix $\mathbf{A}$ that is functionally equivalent. We dub it the echo generation operator, which is a corresponding adjoint operator of the imaging operator $f_{ig}(\mathcal{Y})$. Specifically, they adhere to the relationship as follows.

$$f_{eg}(\mathcal{X}) \approx \mathcal{Y},\ f_{ig}(\mathcal{Y}) = f_{eg}^{\dagger}(\mathcal{Y}) \approx \mathcal{X} \qquad (4)$$

Where $f_{eg}^{\dagger}$ denotes the adjoint operator. The imaging operator is a combination of steps originating from the matched-filtering imaging method. Taking the RMA as an example.

$$f_{ig}(\mathcal{Y}) = \text{ifft}_3\left(\text{fft}_3(\mathcal{Y}) \odot \mathcal{P}_{\mathcal{C}}\right) \qquad (5)$$

Where $\text{fft}_3 / \text{ifft}_3$ denotes two-dimensional DFT/IDFT on each frontal slice $\mathcal{Y}(:,:,i)$ [23], along the range direction, and $\mathcal{P}_{\mathcal{C}}$ denotes the filter tensor that compensates the corresponding phase [6]. $\odot$ denotes the point-wise Hadamard product. Then, the echo generation operator can be derived through their corresponding inverse steps as follows.

$$f_{eg}(\mathcal{X}) = \text{ifft}_3\left(\text{fft}_3(\mathcal{X}) \odot \mathcal{P}_{\mathcal{C}}^*\right) \qquad (6)$$

Where $\mathcal{P}_{\mathcal{C}}^*$ is the conjugate of $\mathcal{P}_{\mathcal{C}}$.

To address the third issue, we introduce the variable splitting technique [17], where alternative variables are utilized to replace the original ones in an alternating direction method of multipliers (ADMM) framework. Specifically, it can be formed as follows.

$$\langle \mathcal{X}, \mathcal{Z}_i, \mathcal{D}_i \rangle = \arg\min_{\langle \mathcal{X}, \mathcal{Z}_i, \mathcal{D}_i \rangle} \frac{1}{2} \|\mathcal{Y} - f_{eg}(\mathcal{X})\|_F^2$$
$$+ \sum_i \left( \beta_i g_i(\mathcal{Z}_i) + \frac{\gamma}{2} \left\| \mathcal{X} - \mathcal{Z}_i + \frac{1}{\gamma} \mathcal{D}_i \right\|_F^2 \right) \qquad (7)$$

Where $\gamma$ is the penalty parameter, and $\mathcal{D}_i$ is the multiplier. And (7) can be solved iteratively. Within the $k$th iteration block, the process contains steps as follows.

$$\mathcal{X}^{k+1} = \arg\min_{\mathcal{X}} \frac{1}{2} \|\mathcal{Y} - f_{eg}(\mathcal{X})\|_F^2 + \frac{\gamma}{2} \|\mathcal{X} - \mathcal{X}_n^k\|_F^2$$
$$\mathcal{Z}_i^{k+1} = \arg\min_{\mathcal{Z}_i} \frac{1}{2} \|\mathcal{Z}_i - \mathcal{Z}_{in}^k\|_F^2 + \frac{\beta_i}{\gamma} g_i(\mathcal{Z}_i) \qquad (8)$$
$$\mathcal{D}_i^{k+1} = \frac{1}{\gamma} D_i^k + \mathcal{X}^{k+1} - \mathcal{Z}_i^{k+1}$$

$\mathcal{X}_n^k = \mathcal{Z}_i^k + (1/\gamma)\mathcal{D}_i^k$ and $\mathcal{Z}_{in}^k = \mathcal{X}^{k+1} + (1/\gamma)\mathcal{D}_i^k$. The common consistency regularization for different tasks is fulfilled first. It can be solved by letting the first-order derivative be zero, as follows.

$$\mathcal{X}^{k+1} = 1/(1+\gamma)\left[\mathcal{X}_n^k + f_{ig}(\mathcal{Y})\right] \qquad (9)$$

Following it, different demands are individually fulfilled through corresponding cognition regularizations. Notably, they are all decoupled. Thus, although various tasks have different cognition regularizations, they can all be fitted into this process. These cognition regularization subproblems have a common feature. In the image space, the input image $\mathcal{Z}_{in}^k$ can be viewed as deviating from the actual location of $\mathcal{Z}$ to a nearby location [19]. Thus, these subproblems can be solved by moving the deviated image back through off-the-shelf proximal operators with the gradient descend [19].

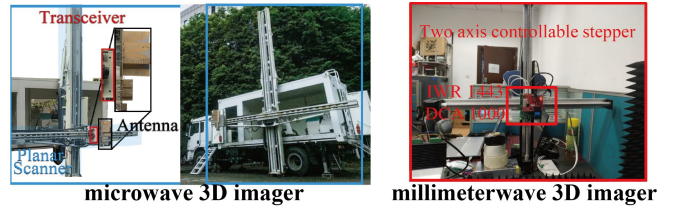

**Fig.3.** Experiment systems. The left one is for task 1, and the right one is for task 2 and task 3.

**Table 1** Experiment parameters

| | Center frequency | Bandwidth | Array size | Center range |
|---|---|---|---|---|
| Task 1 | 11 GHz | 2 GHz | 5 m × 5 m | ≈ 15 m |
| Task 2 | 79 GHz | 4 GHz | 0.4 m × 0.4 m | ≈ 0.6 m |
| Task 3 | 79 GHz | 4 GHz | 0.4 m × 0.4 m | ≈ 0.6 m |

Taking the formerly mentioned tasks as examples. 1) *Scattering diagnosis*. The first demand can be fulfilled by the target's $l_p$ norm regularization with the threshold-based function [24] as the proximal operator. And the second one can utilize the noise-and-clutter's $l_2$ norm regularization [25], with its operator being the shrinkage function [19]. 2) *Person screen imaging*. Without the high precision requirement, the first demand can utilize a less complex $l_1$ norm with a shrinkage-threshold function [10] proximal operator. And the second demand can be met through Shearlet-Transform [26] which is capable of characterizing directional distributed-scatters. And the regularization term can be the $l_1$ norm in the transformed domain. 3) *Parcel screen imaging*. Apart from the above two, the kernel norm of interferences helps suppress interferences from surroundings, which present dense and redundant patterns [27] shown in the experiments, the kernel norm of interferences is helpful. And the operator is the shrinkage-threshold function on its singular values [23].

## 3. EXPERIMENTS

Experiments are conducted on data from two 3D radar imaging systems for the three tasks mentioned earlier. The main parameters are listed in Tab.1, and the system's photos are shown in Fig.3.

### 3.1. Task 1: scattering diagnosis task

The scene shown in Fig.4(a) contains three scatters of different radar cross sections (RCS). Due to limited pages, we just present 2D projected top-view images in this one. The sparse-oriented is based on $l_1$ norm. We can see that compared with the MF result in Fig.4(b), ours in Fig.4(d) has less clutter and noise, and the resolution is higher. And compared with the sparse-oriented (SRO) result in Fig.4(c), ours has higher energy precision. This is due to the task demands being taken into consideration, with the $l_q$ norm to avoid biased estimation. Qualitatively, the relative energy errors of three scatters are calculated for the last two. The results of SRO are 8.4%, 14.2%, and 37.3%, respectively. While for the proposed one, they are 2.9%, 3.8%, and 1.2%, respectively. The average error is 2.6%, with an increment of 17.4%.

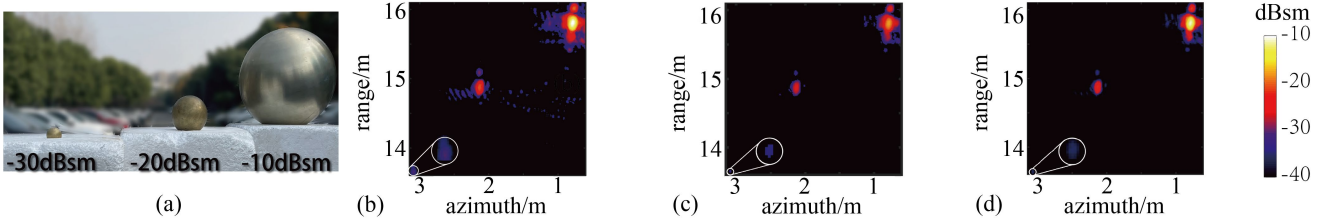

**Fig.4.** Results of task 1. (a) Target photo. (b-d) Results of MF, SRO, and the proposed, respectively. Pixel color: calibrated RCS value.

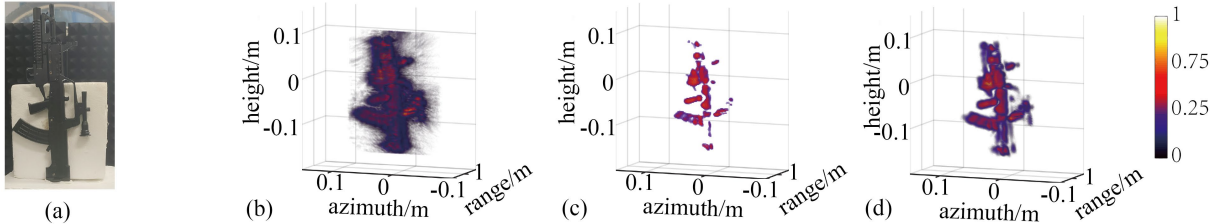

**Fig.5.** Results of task 2. (a) Target photo. (b-d) Results of MF, SRO, and the proposed, respectively. Pixel color: relative amplitude.

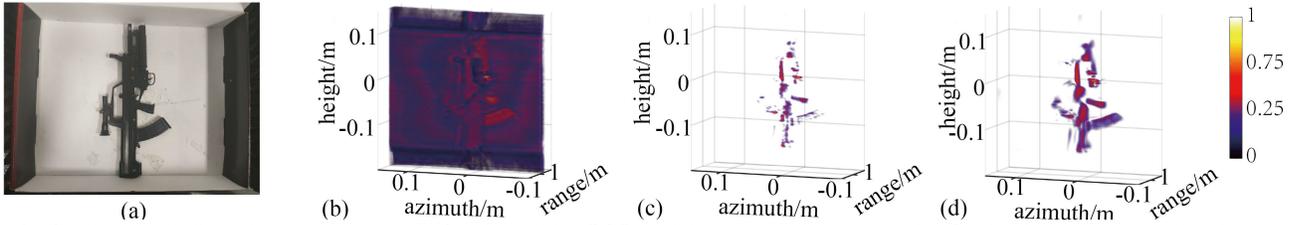

**Fig.6.** Results of task 3. (a) Target photo. (b-d) Results of MF, SRO, and the proposed, respectively. Pixel color: relative amplitude.

### 3.2. Task 2: person screen imaging

The scene contains a rifle model in the free space, as shown in Fig.5(a). Here, we consider that the person's cloth is thin, which has little effect on the target's imaging result. Thus, we just put the rifle in the free space instead of on a human body model [7]. Obviously, in Fig.5(b), lots of clutter-and-noise exist in the result of MF. And the shape information of SRO's result in Fig.5(c) is extremely lost. Ours in Fig.5(d) has the best visual performance that makes a good balance between the former two ones. Qualitatively, we calculate the SSIM metric [13] to evaluate this shape information loss. The reference image is the enhanced MF image through post-processing with both hand-crafted enhancements and the target-extraction algorithm [28]. The results are 0.51 (MF), 0.79 (SRO), and 0.95 (proposed). The higher the score is, the more the shape information is retrieved. Ours achieves the highest score, which matches the visual perception.

### 3.3. Task 3: parcel screen imaging task

The scene contains a rifle model in a box, as shown in Fig.6(a). Here, the box's influence can no longer be ignored. As seen from the result of MF in Fig.6(b), the rifle's image is covered by the box's interference image. As we described before, it has a highly redundant and dense pattern with little texture. Mathematically, it's low-rank. Thus, we use the kernel norm to characterize. The result of the proposed framework in Fig.6(d) suppresses the interference and still maintains the shape to a certain degree. The result of SRO in Fig.6(c) is much worse. As the target and the interference are mixed, to suppress the interferences, only a few strong scatters are retrieved, which would not be helpful to target detection and recognition. However, compared to the last experiment, structure loss exists in our result due to the interference. Qualitatively, we calculate the target-to-background ratio (TBR) metric [10] to evaluate the suppression performance. The results are 19.5dB (MF) and 56.2dB (proposed), with a 36.7dB suppression gain. Above three experiments verify the effectiveness. Ours shows the flexibility that can handle various tasks. Regarding the task-oriented information retrieval, ours achieve spurious performance.

## 4. CONCLUSION

Faced with the 3D radar imaging inverse problems, we propose a multi-cognition task-oriented imaging framework. As more information can be retrieved with the improvement of the wide-bandwidth radar system, the aim is to enable task-depended information retrieval, which hasn't been studied quietly in the area. Unlike the current ones that only differentiate the target and the background aspects, we take a step further by analyzing and differentiating different task demands. Then these demands are fulfilled through a multi-cognition regularization model. And a generalized imaging method to solve it is proposed with the help of approximation and variable-splitting techniques. Experiments on different tasks verify the effectiveness of the framework. Currently, the solving of the model is still by traditional optimization solvers. As more cognition regularizations are embedded, the optimal hyper-parameters they need may be inconvenient to find. An alternative and promising way may be the recently proposed Learning to Optimize (L2O) method, where deep learning is utilized to unfold the optimization process [29]. We have validated its effectiveness for the only one cognition regularization [13], [30] case. The multiple-cognition scenario would be our next-step work. For the last two screen imaging tasks, we have collected more data and revealed them recently [6].